\documentclass[a4paper,fleqn,usenatbib]{mnras}

\usepackage{graphics,epsf}
\usepackage{amsmath}                % American Mathematical Society package
\usepackage{amsfonts}               % American Mathematical Society fonts
\usepackage{amssymb}                % American Mathematical Society symbol
\usepackage{epsfig}                 % EPS figures
\usepackage{graphicx}
\usepackage{xcolor}

\definecolor{redak}{rgb}{0.9,0.15,0.05}
\def \msyr{\rm{M_{\odot}}~\rm{yr^{-1}}}
\def \cm{~\rm{cm}}
\def \s{~\rm{s}}
\def \km{~\rm{km}}

\def \erg{~\rm{erg}}

\def \yr{~\rm{yr}}

\def \days{~\rm{day}}

\def \rmModot{~\rm{M_\odot}}
\def \rmRodot{~\rm{R_\odot}}
\def \rmLodot{~\rm{L_\odot}}

\title[ILOTs by Binary Interaction and Jets]{Explaining Two Recent Intermediate Luminosity Optical Transients (ILOTs) by a Binary Interaction and Jets}

\author[N. Soker and A. Kashi]{
Noam Soker$^{1}$\thanks{E-mail: \href{mailto:soker@physics.technion.ac.il}{soker@physics.technion.ac.il}}
and
Amit Kashi$^{1}$\thanks{E-mail: \href{kashia@physics.technion.ac.il}{kashia@physics.technion.ac.il}}
\\
$^{1}$Physics Department, Technion -- Israel Institute of Technology, Technion City -- Haifa 3200003, Israel\\\
}

\date{Accepted XXX. Received YYY; in original form ZZZ}

\pubyear{2016}

\begin{document}
\label{firstpage}
\pagerange{\pageref{firstpage}--\pageref{lastpage}}
\maketitle

\begin{abstract}
We propose that two recent intermediate luminosity optical transients (ILOTs), M31LRN 2015 and SN 2015bh (SNHunt 275; PTF 13efv) can be accounted for with a stellar binary model involving mass transfer that leads to the launching of jets.
We inspect observations of the ILOT M31LRN 2015 and conclude that it cannot be explained by the onset of a common envelope evolution (CEE).
Instead we conjecture that a $M \simeq 1 \textrm{--} 3 \rmModot$ main sequence star accreted $\simeq 0.04 \rmModot$ from the giant star, possibly during a periastron passage.
The main sequence star accreted mass through an accretion disc, that launches jets.
The radiation from the disk and the collision of the jets with the ambient gas can account for the luminosity of the event.
Along similar lines, we suggest that the 2013 eruption of SN 2015bh (SNHunt~275) can also be explained by the High-Accretion-Powered ILOT (HAPI) model.
In this case a massive secondary star $M_2  \gtrsim 10 \rmModot$ accreted $\approx 0.05 \rmModot$ from a much more massive and more evolved star during a periastron passage.
If the much more energetic 2015 outburst of SN 2015bh (SNHunt 275) was not a supernova explosion, it might have been a full almost head-on merger event, or else can be accounted for by a the HAPI-jets model in a very highly eccentric orbit.
\end{abstract}


% ==========================================================
\section{INTRODUCTION}
 \label{sec:intro}
% ==========================================================

Eruptive stars with peak luminosity values between the typical luminosities of novae and supernovae (SN) form an heterogeneous group (e.g. \citealt{Mouldetal1990, Rau2007, Ofek2008, Prieto2009, Botticella2009, Smithetal2009, Berger2009b, KulkarniKasliwal2009, Mason2010, Pastorello2010, Kasliwaletal2011, Kasliwal2011, Tylendaetal2013, Kasliwal2013}). Some of these objects are low luminosity SNe and related objects, such as Ca-rich transients and .Ia SNe, many of which are powered by thermonuclear outbursts and explosions.

The remaining gap objects that are not supernovae are part of a still heterogeneous group that is generally termed Intermediate Luminosity Optical Transients (ILOTs; \citealt{Berger2009b, KashiSoker2016}).
\cite{KashiSoker2016} further classified ILOTs into three types of objects:
\begin{enumerate}
\item Intermediate-Luminous Red Transients (ILRT). These are ILOTs of evolved stars, such as asymptotic giant branch (AGB) or extreme-AGB (ExAGB) stars, like NGC~300~OT2008-1 (NGC~300OT; \citealt{Monard2008, Bond2009, Berger2009b}) and SN~2008S \citep{ArbourBoles2008}.
\item Giant eruptions of luminous blue variables (LBV) and SN Impostors. Examples include the Great Eruption (GE) of $\eta$ Carinae in the years 1837--1856, and the pre-explosion eruptions of SN~2009ip. Within the context of the binary model LBV giant eruptions might be considered in some sense to be the massive relatives of ILRTs \citep{KashiSoker2016}.
\item Luminous Red Novae (LRN) or Red Transients (RT) or Merger-Bursts. These outbursts are powered by a full merger of two stars. The process of destruction of the less dense star, on to the denser
star or inside its envelope, releases gravitational energy that powers the transient. Examples include V838 Mon and V1309~Sco.
Merger events of stars with sub-stellar objects are also included.
\end{enumerate}
As more ILOTs are being discovered we add them to the Energy-Time Diagram\footnote{An updated version of the ETD is available at \url{http://physics.technion.ac.il/~ILOT/}}
(ETD), which shows their total energy against the eruption duration \citep{KashiSoker2016}.
Many of the ILOTs sit on the Optical Transient Stripe (OTS) in the ETD, suggesting they are powered by a similar source of energy.

Models for ILOTs include single-star models (e.g., \citealt{Thompsonetal2009, Kochanek2011} for ILRTs and \citealt{Ofeketal2013} for a SN impostor),
and binary stellar models (e.g., \citealt{Kashietal2010, KashiSoker2010b, SokerKashi2011, SokerKashi2012, SokerKashi2013, McleySoker2014}).
We note that \cite{Adamsetal2015} cast doubt that the progenitor star of the ILRTs NGC~300OT and SN~2008S survived the ILRT event.
In a merger processes the secondary star can survive the first encounter, in which case the two stars form a common envelope (CE), or alternatively be destroyed on encounter.
The close binary interaction can lead to enhanced mass loss and mass transfer.
\cite{Pejchaetal2016a} and \cite{Pejchaetal2016b} based their model for ILOTs on a high rate of mass loss through the second Lagrangian point.
The collision of equatorially ejected mass transfers kinetic energy to radiation. No jets are considered in their model.

A mass transfer process, whether in close detached systems, or in the grazing envelope evolution (GEE), or during the CE evolution (CEE),
can lead to the formation of an accretion disc or an accretion belt around the more compact star. Such a disc might launch jets.
The powering of ILOTs by accretion onto the more compact star in a binary system is termed the High-Accretion-Powered ILOT (HAPI) model,
and was developed by us in an earlier paper \citep{KashiSoker2010b}.
The HAPI model was then applied to the formation of ILOT events during the GEE \citep{Soker2016NewA}.
In some cases the interacting system is in fact a triple stellar system.
The tertiary star might induce orbital instabilities and causes the two inner stars to interact, and in other cases all three stars can participate in the mass transfer process.

Several ILOTs have been attributed to the CEE of a binary stellar system, or to the onset of the CE.
\cite{SokerTylenda2003} and \cite{TylendaSoker2006} attributed the ILOT V838~Mon to a merger process of two stars where the low mass star had been destroyed. \cite{RetterMarom2003} and \cite{Retteretal2006}, on the other hand, suggested that V838~Mon was powered when planets entered the envelope of the star and formed a CE.
Scenarios of CEE with a stellar companion followed with the ILOTs OGLE-2002-BLG-360 \citep{Tylendaetal2013}, {V1309~Sco} \citep{Tylendaetal2011, Ivanovaetal2013a, Nandezetal2014, Kaminskietal2015}, and recently M31LRN~2015 \citep{MacLeodetal2016}.
In section \ref{sec:M31LRN} we study the CEE scenario that was proposed by \cite{MacLeodetal2016} to account for M31LRN~2015 and propose the HAPI
model as an alternative explanation for the same ILOT.

In some cases the two jets that are launched from the compact companion might expel more mass from the system, and form an expanding bipolar nebula \citep{KashiSoker2010a}, such as the
bipolar nebula of $\eta$ Carinae, the Homunculus, that was formed in
the GE (e.g., \citealt{HumphreysMartin2012}).
$\eta$ Carinae is known to be a binary systems \citep{Damineli1996} that did not enter a CEE.
The sharp peaks in the light curve during the GE occurred around periastron passages of the binary system \citep{Damineli1996, KashiSoker2010a, SmithFrew2011}.
\cite{SokerKashi2013} speculated that the progenitor of SN~2009ip was in a binary system, and suggested that the pre-explosion outbursts of SN~2009ip occurred
during, and as a result of, periastron passages.

In section \ref{sec:PTF13efv} we analyze the ILOT SN~2015bh (aka SNHunt~275; PTF~13efv; PSN J09093496+3307204) that had at least two outbursts.
It is not clear yet whether the last peak was casued by a real SN (e.g., \citealt{Postigoetal2015}) or an impostor.
Several works (e.g., \citealt{EliasRosaetal2015, EliasRosaetal2016, RichardsonArtigau2015, Thoneetal2016}) noticed that SN~2015bh has some similarities with SN~2009ip.
\cite{Ofeketal2016} analyzed the behavior of SN~2015bh and discussed its behavior within the context of a single star suffering an outburst with a super-Eddington luminosity. We instead propose a binary model.

In section \ref{sec:summary} we summarize by concluding that a binary model based on jets, the HAPI-jets model, seems to explain the best many of the properties of ILOTs.

% ==========================================================
\section{THE ILOT M31LRN~2015}
 \label{sec:M31LRN}
% ==========================================================

The ILOT M31LRN~2015 was discovered in January 2015 \citep{Shumkovetal2015}, was compared to the merger-burst (LRN) V838~Mon (e.g., \citealt{Kurtenkovetal2015}), and was suggested to be a result of a merger process (e.g., \citealt{Dongetal2015, Williamsetal2015}), i.e., be an LRN (RT; or merger-bursts).
The bolometric light curve can be divided into two parts.
A rise to the peak and decline, lasting from about -10 days to +10 days relative to the peak, and a plateau phase of about constant luminosity lasting for about another 40 days.

% ================================
\subsection{A merger-burst model}
 \label{subsec:mergerburst}
% ================================

In a recent paper \cite{MacLeodetal2016} propose a scenario for M31LRN~2015 where a main-sequence (MS) secondary star of mass $M_2 = 0.1 \textrm{--} 0.6 \rmModot$ entered a CEE with a giant primary star of mass $M_1 =3 \textrm{--} 5.5 \rmModot$ and a radius of $R_1 \simeq 35 \rmRodot$.
The process that leads to the CEE, according to their model, is the Darwin instability.
They further suggest that the main energy source of the radiated energy of $5 \times 10^{45} \erg$ during the plateau phase, about 10 to 50 days past the peak, is the recombination of the ejected mass.
According to their model, this requires that the ejected mass that recombined amounts to at least $\Delta m_{\rm ej,plateau}= 0.17 \rmModot$.
We see several problems in the model proposed by \cite{MacLeodetal2016}, as we explain in the following subsections. We then propose an alternative scenario.

% ================================
\subsubsection{Energy considerations}
 \label{subsubsec:energy}
% ================================
ILOTs that are powered by merger (merger-bursts) need not be powered by accretion of mass onto the compact companion. Such is V1390~Sco \citep{Tylendaetal2011}. However, there are significant differences between V1390~Sco and the model \cite{MacLeodetal2016} propose for M31LRN~2015.
The binary orbital period of the progenitor of V1309~Sco was 1.4 day, with a radius of the primary star of $3 \textrm{--} 5 \rmRodot$ \citep{Tylendaetal2011}. The radiated energy in the outburst of V1390~Sco was $E_{\rm rad} \approx 3 \times 10^{44} \erg$ \citep{Tylendaetal2011}. The energy stored in the orbital motion of the binary system, $E_{\rm orb} \approx 0.2 \textrm{--} 5.6 \times 10^{47} \erg$, was about two orders of magnitude higher than the radiated energy during the outburst \citep{Tylendaetal2011}
\begin{equation}
Q_r({\rm V1390~Sco}) \equiv \frac{E_{\rm orb}}{E_{\rm rad}} \approx 70 \textrm{--} 2000.
\label{eq:Qrsco}
\end{equation}
The energy stored in the orbital motion of the progenitor of V1390~Sco can easily account also for the kinetic energy of the ejected matter which is about $10$ times as large as the radiated energy, and for the energy required to inflate the envelope \citep{Tylendaetal2011}.

In the model of \cite{MacLeodetal2016} for M31LRN~2015 the energy stored in the orbital motion is $E_{\rm orb} \approx 0.2 \textrm{--} 1.8 \times 10^{47} \erg$.
The radiated energy during the peak and plateau combined is $E_{\rm rad} \approx 8 \times 10^{45} \erg$, and we find
\begin{equation}
Q_r({\rm M31LRN}) \approx 2 \textrm{--} 22.
\label{eq:Qrm31}
\end{equation}
The typical value of $Q_r$ for V1309~Sco is $\approx~50$ times larger than what the binary progenitor model proposed by \cite{MacLeodetal2016} gives for M31LRN~2015.
We note that the relevant energy to consider is the energy released by the binary system as the secondary stars spirals-in deep to the giant star. This is done in section \ref{subsubsec:mass}.
Furthermore, neither the scenario proposed by \cite{MacLeodetal2016} nor the one we propose attribute a specific role to the value of $Q_r$.
Despite these two limitations, the large ratio $Q_r(V1309 Sco) / Q_r(M31LRN) \approx 50$ suggests that V1390~Sco cannot be used to characterize M31LRN,
or to conclude that it was powered by the onset of a CEE.
The particular binary parameters adopted in the model of \cite{MacLeodetal2016} are weakly constrained by observations,
and the energy range they provide is quite large.
Only if the maximum value applies we can get $Q_r({\rm V1390~Sco})\approx3Q_r({\rm M31LRN})$. This is very unlikely.

The quantity $E_{\rm orb}$ is the relevant energy if the secondary star motion is slowed down rapidly in the outer regions of the primary star.
However, the large radius of the progenitor of M31LRN~2015 of $R_1 \simeq 35 \rmRodot$ \citep{MacLeodetal2016} implies that the density in its envelope is very low, and the much denser MS companion will not slow down much in the outer envelope. The secondary star must penetrate deep into the envelope.

Based on crude energy considerations alone, namely that $Q_r({\rm M31LRN}) \ll Q_r({\rm V1390~Sco})$, we conclude that it is unlikely that the merger-burst model that applies to V1309~Sco can be scaled to explain the outburst of M31LRN~2015.

% ================================
\subsubsection{Ejected mass}
 \label{subsubsec:mass}
% ================================

For about 30 days, from the peak to 30 days post-peak, the photosphere of the ejected material expands with a velocity of $v_{\rm ej} \simeq 400 \km \s^{-1}$ \citep{MacLeodetal2016}.
As according to their model the radiated energy is recombination energy, the recombining gas in the first 30 days must move at about this velocity.
The recombining gas in the last 20 days can move at a lower velocity of $\approx 300 \km \s^{-1}$.
Overall, the kinetic energy of the ejected gas in their model is $E_{\rm k,ej} \simeq 2 \times 10^{47} \erg$.

The most massive secondary in their model has a mass of $M_2=0.6 \rmModot$.
The secondary star has to spiral-in to a radius of $r_{\rm 2,ej}$ to account for the kinetic energy,
given by
\begin{equation}
E_{\rm k,ej} \lesssim \frac {G M_1(r_{\rm2,ej}) M_2}{2r_{\rm2,ej}} - \frac {G M_1 M_2}{2R_1}.
\label{eq:r2ej1}
\end{equation}
In the above equation part of the energy on the right-hand-side must go to radiation, and for that we used the `$\lesssim$' sign.
However, as mentioned above, the radiated energy is much smaller than the kinetic energy so we can neglect it.
Taking $M_1=5\rmModot$, $M_2=0.6 \rmModot$ and $R_1=35 \rmRodot$ gives
that the final orbital energy of the binary system at the final orbit of the secondary star is
\begin{equation}
E_{\rm 2,f} \equiv E_{\rm k,ej} + \frac{G M_1 M_2}{2R_1} \simeq 3.6 \times 10^{47} \erg.
\label{eq:E2f}
\end{equation}

\begin{figure}
\centering
\includegraphics[trim= 0.2cm 0.15cm 0.9cm 1.2cm,clip=true,width=0.99\columnwidth]{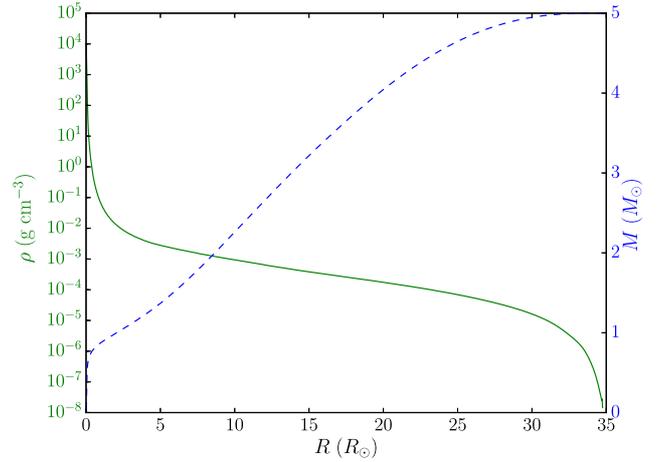}
%trim=l b r t
\caption{
A \texttt{MESA} model for a $5\rmModot$ AGB star. The effective temperature of the model is $T_{\rm{eff}} = 4\,590 \rm{K}$,
the stellar radius is $R_1 = 34.8 \rmRodot$, and the luminosity is $L_1 = 482 \rmLodot$.
$M$ is the total mass inwards to $R$.
}
\label{fig:profile4msun}
\end{figure}

With the goal to asses the mass enclosed in each radius, we run a model of a $M_{\rm ZAMS} = 5\rmModot$ star using \texttt{MESA} \citep{Paxtonetal2011} and let it evolve to the AGB stage, to the point when its radius is
$R_1 \simeq 35 \rmRodot$, as estimated by \cite{MacLeodetal2016}.
Our model is shown in Figure \ref{fig:profile4msun}.
% FFFFFFFFFFFFFF
% FFFFFFFFFFFFFF

To satisfy equation (\ref{eq:E2f}) the secondary star must spiral-in to a radius of
\begin{equation}
r_{\rm2,ej} \approx 3.8  %3.813
\left[ \frac{M_1(r_{\rm2,ej})}{1.2 \rmModot} \right]   %5-3.793
\left( \frac{M_2}{0.6 \rmModot} \right)
\left( \frac{E_{\rm 2,f}}{3.6 \times 10^{47} \erg} \right)^{-1} \rmRodot ,    %3.626e+47
\label{eq:r2ej2}
\end{equation}
where the radius $r_{\rm2,ej}$ and mass $M_1(r_{\rm2,ej})$ are scaled according to the solution of equation (\ref{eq:r2ej2}) with the model presented in Figure \ref{fig:profile4msun}.

The mass of the primary star that resides above radius $r_{\rm2,ej} \simeq 3.8 \rmRodot$ is
$M_1(R_1) - M_1(r_{\rm2,ej}) \simeq 3.8 \rmModot$.
If we take half the above value of kinetic energy, $E_{\rm k,ej} \simeq 10^{47} \erg$, we get $r_{\rm2,ej} \simeq 8.7 \rmRodot$ and the mass above it is
$M_1(R_1) - M_1(r_{\rm2,ej})\simeq 2 \rmModot$.
Even if we consider that not all this mass was ejected, these values are still much larger than the ejected mass of $\Delta m_{\rm ej,plateau} = 0.17 \rmModot$ according to the model of \cite{MacLeodetal2016}.
Namely, the secondary star deposits its orbital energy to a mass much larger than a $0.17 \rmModot$.
So for the $0.17 \rmModot$ to acquire a kinetic energy of $E_{\rm k,ej} \simeq 1 \textrm{--} 2 \times 10^{47} \erg$, the secondary must release more gravitational energy.
This implies that it spirals-in deeper than $r_{\rm2,ej}$  estimated above.
It is not clear at all that under these conditions, where the energy is distributed among several solar masses, a small mass of $\Delta m_{\rm ej,plateau} = 0.17 \rmModot$
can escape with a velocity that is about twice the escape velocity from the binary system.

% ================================
\subsubsection{Time scale}
 \label{subsubsec:timescale}
% ================================

Another problem we see in in the model proposed by \cite{MacLeodetal2016} concerns the time scale of envelope ejection.
In the case of a powering by recombination the photosphere moves inward in the mass coordinate.
Hence, most of the ejected mass was ejected at the same time and at the same velocity of $v_{\rm ej} \simeq 400 \km \s^{-1}$.
However, as was shown in section \ref{subsubsec:mass} the secondary star needs to spiral-in to a very small radius. This requires at least one dynamical time at the surface, and likely much more.
Namely, the ejection time will last over a time longer than 10 days. This is a substantial fraction of the 30 days during which the photosphere expands with a constant velocity.  This does not fit the observations, unless later ejecta are moving at higher velocities than $v_{\rm ej} \simeq 400 \km \s^{-1}$ and catch-up with the photosphere.
But this makes the energy and mass problems discussed in the previous subsections even more severe.

% ================================
\subsection{An alternative ILRT model}
 \label{subsec:ILRT}
% ================================

We propose that the ILOT M31LRN~2015 was not powered from the merger process itself, but rather by accretion onto a companion. Namely, instead of a merger-burst model (or LRN or RT; see section \ref{sec:intro} for terminology), it was powered by a companion accreting mass from a giant star, namely, an ILRT. As for powering the radiation, instead of recombination energy we suggested that the accreted energy is channelled to radiation and kinetic energy of jets (winds).
The collision of jets with previously ejected mass can convert more kinetic energy to radiation.
The system is more like an ILOT during a grazing envelope evolution \citep{Soker2016NewA}.

\cite{MacLeodetal2016} estimate that the kinetic energy is 4--17 times the radiated energy, depends on the velocity.
We shall adopt a total energy of $E_{\rm tot} \approx E_{\rm k,ej} \approx 10^{47} \erg$,
but note that the energy can be even smaller, as our model is not based on recombination, and can work with ejected mass of much less than $0.17 \rmModot$.
We scale the mass and radius of the companion with $M_2=2 \rmModot$ and $R_2=1.5 \rmRodot$, respectively.
If the energy comes from accretion onto the companion, then in order to supply the total energy the accreted mass should be
\begin{equation}
M_{\rm{acc}} \simeq 0.04 \left(\frac{E_{\rm tot}}{10^{47} \erg} \right) \left( \frac{M_2}{2 \rmModot} \right)^{-1} \left( \frac{R_2}{1.5 \rmRodot} \right) \rmModot.  %0.0395
\label{eq:macc}
\end{equation}
The escape velocity from the companion is $v_{\rm esc} \simeq 700 \km \s^{-1}$, and so it can easily account for ejecta speed of $v_{\rm ej} \simeq 400 \km \s^{-1}$.   %713

The accretion rate implied by equation \ref{eq:macc} for an event duration of $0.1 \yr$, is $\approx 0.4 \msyr$.
This accretion rate is extremely high.
\cite{Shiberetal2016} constructed a scenario by which a solar type star can accrete mass at a rate of
$\approx 0.01 M_\odot \yr^{-1}$, and more massive stars can accrete at an even high rate.
To accrete at such a high rate the accreted gas must get rid of its extra energy. Most of it carried in the jets \citep{Shiberetal2016}.
To account for the radiated energy, it is required that about 10 per cent of the kinetic energy carried by the jets is radiated.

To estimate the time scale for the event in our model, we shall assume that accretion occurs through a disc, and that the time scale of the event is in the order of the viscosity time of the disc $t_{\rm vis}$.
Assuming a simple thin $\alpha$-disc, the viscosity is $\nu = \alpha c_s H$, where $c_s$ is the spped of sound  and $H$ is the disc hight (\citealt{ShakuraSunyaev1973}),
we get
\begin{equation}
\begin{split}
t_{\rm vis} &\approx \frac{R^2}{\nu} = \frac{R^2}{\alpha c_s H} \approx \frac{t_{\rm Kep}}{2 \pi \alpha} \left(\frac{R}{H}\right)^2 \\
&\approx 24 \left(\frac{\alpha \vphantom{X}}{0.1 \vphantom{X}}\right) \left(\frac{R}{10H}\right)^2 \left( \frac{M_2}{2 \rmModot} \right)^{-1/2} \left( \frac{R_2}{1.5 \rmRodot} \right)^{3/2} \rm{days}.  %23.96
\end{split}
\label{eq:tvis}
\end{equation}
In the above equation we took the value of the Keplerian time on the stellar surface $t_{\rm Kep} \simeq 3.6 ~\rm{h}$. %16350sec
On the other hand at these very high accretion rates the disk is likely to be thicker \citep{Shiberetal2016}, and the term $(R/H)^2$
can reduce the time scale by about an order of magnitude, i.e., $t_{\rm vis} \approx \rm{several} \times$ day.
As the disc extends to somewhat larger than the secondary surface, the viscosity time is estimated to be $\approx 1 \textrm{--} 3$~weeks.
Overall we get that $t_{\rm vis}$ is about the time scale of the observed event.

% ==========================================================
\section{THE ILOT SN~2015bh (SNH\lowercase{unt}~275)}
 \label{sec:PTF13efv}
% ==========================================================

% ================================
\subsection{The single-star super-Eddington model}
 \label{subsec:superEdd}
% ================================

The ILOT SN~2015bh (SNHunt~275) holds several puzzles (e.g., \citealt{Postigoetal2015, EliasRosaetal2015, EliasRosaetal2016, RichardsonArtigau2015, Ofeketal2016, Thoneetal2016}).
It undergone a strong outburst in 2015, a weaker one in 2013, and possibly an earlier one in 2009.
First and most important is whether the last outburst was a terminal SN explosion.
\cite{Ofeketal2016} bring arguments that might suggest that it was not a SN explosion. We here accept this view, and examine its consequences. We note though that \cite{EliasRosaetal2016} argue that the last outburst was a faint SN explosion.
The second puzzle is whether the detection on 2009 Sep 10 that looks like an outburst is real. \cite{Ofeketal2016} analyzed it and conclude that it is not an outburst, but rather it is likely a bad pixel or radiation hit event (i.e., cosmic ray).
We accept this conclusion, despite three interesting coincidences that otherwise might have hinted at a real detection.
\begin{enumerate}
\item The luminosity of the one-point 2009 peak (MJD 55084.5089) was only $\approx 7$ per cent above the maximum one in the 2013 outburst.
\item The time of the detection in 2009 took place $\approx 2250 \textrm{--} 2070 \days$ before $t_0=2457157.36$ (the time of the eruption in 2015).
This is about 4 times the interval of $\approx 530 \days$ between the 2013 and 2015 peaks.
\item The one point detection before and one point detection after the 2009 peak are more luminous than the others points outside the peak in 2009.
\end{enumerate}
The new light-curve presented by \cite{Thoneetal2016} indicate that there is high emission earlier in 2009. The light-curve also presents an earlier outburst in 2008, and a general complicated behavior. With the present data we refrain from fitting an orbital period for the system, and leave this question open.

\cite{Ofeketal2016} discuss the behavior of SN~2015bh within the context of a single star suffering an outburst with a super-Eddington luminosity
following the model proposed by \cite{Shaviv2000, Shaviv2001}.
We calculate the optical depth of the super-Eddington outburst model of \cite{Ofeketal2016}.
According to their results, the mass lost in the 2013 eruption is $M \approx 4\times 10^{-5} \rmModot$.
Accounting for $\Delta t \simeq 20 \days$ the average mass loss rate is $\dot{M} \approx 7\times 10^{-4} \msyr$. %7.3e-4
Taking their observed ejecta velocity $v=1000 \km \s^{-1}$, their photospheric radius\footnote{
\cite{Ofeketal2016} state a maximum photosphere radius (excluding the region
responsible for the Balmer lines) of $4 \times 10^{14} \cm$.
Following a comment by N. Shaviv (private communication) we take the photosphere to be
at $1.4 \times 10^{14} \cm$.}
$r=1.4\times 10^{14} \rm{cm}$ ,opacity of $\kappa=0.4$, and assuming spherical symmetry, the optical depth for a wind blown for a very long time is
\begin{equation}
\begin{split}
\tau_{2013} &\approx 0.1
\left( \frac{\kappa}{0.4~\rm{cm^2~g^{-1}}} \right)
\left( \frac{\dot{M}}{7\times 10^{-4} ~ \msyr} \right) \\
&
\left( \frac{r}{1.4\times 10^{14} ~ \rm{cm}} \right)^{-1}
\left( \frac{v}{\vphantom{\times}10^3 \km \s^{-1}} \right)^{-1}.
\end{split}
\label{eq:tau}
\end{equation}
This is too low to account for a photosphere.
We see this as a challenge to the single-star-super-Eddington outburst model.

% ================================
\subsection{An alternative binary model}
 \label{subsec:binarymodel}
% ================================

We propose that instead of a super-Eddington outburst, SN~2015bh is (or was) a massive eccentric binary stellar system that underwent a giant eruption in 2013,
followed by a strong eruption in 2015, which might have been a SN or a stellar merger event.
The binary period might be either $\simeq 530 \days$ or a simple fraction of this number.
This raises the possibility that the mechanism behind the 2013 outburst is similar to the mechanism behind the GE of $\eta$ Carinae.

If the 2015 peak was not the result of a terminal event, i.e., a SN explosion or a stellar merger event, then the
binary model allows for another eruption at the next periastron passage.
If the orbital period is indeed $\simeq 530 \days$, the next event might take place in October 2016.

Applying the HAPI-jets model, the energy of the 2013 eruption, and the 2015 eruption if was not a terminal event (which we consider unlikely),
came from accretion of gas to the binary companion, which lead to the formation of accretion disk and the launching of jets.
A prediction of the model is a bipolar gas-ejection morphology, such as that during the GE of $\eta$ Carinae.

As an illustrative example we assume that the energy results from accretion onto a MS secondary of $M_2= 10 \rmModot$ and radius $R_2= 4 \rmRodot$,
and the radiated energy of the eruption is a fraction $\beta$ of the accretion energy.
The secondary star can be more massive, but the gravitational potential on its surface will not changes much if it is a MS star.
We can then estimate the accreted mass in the 2013 eruption to be
\begin{equation}
\begin{split}
M_{\rm acc,2013} &\approx 0.05
\left( \frac{\beta}{0.1} \right)^{-1}
\left( \frac{E_{\rm rad,2013}}{2.4\times 10^{46} \, \erg} \right) \\
&
\left( \frac{R_2}{4 \rmRodot} \right)
\left( \frac{M_2}{10 \rmModot} \right)^{-1}
\rmModot .
\end{split}
\label{eq:macc2013}
\end{equation}

For an event length of $\approx 0.1 \yr$, the implied accretion rate is $\approx 0.5 \msyr$.
According to the high accretion rate model proposed by \cite{Shiberetal2016} a main sequence star of $\gtrsim 10 \rmModot$ can accommodate such a high accretion rate if launches energetic jets.
These are the jets that collide with previously ejected gas and emit the radiation.
In the process proposed by \cite{Shiberetal2016} the accreting star itself does not radiate much above the Eddington luminosity.
Most of the energy is rather carried by jets, i.e., kinetic energy.
The collision of the jets with the ambient gas heats the gas and leads to the highly super-Eddington luminosity.

If we repeat the estimate for the 2015 eruption, we find
\begin{equation}
\begin{split}
M_{\rm acc,2015} &\approx 4      %3.8
\left( \frac{\beta}{0.1} \right)^{-1}
\left( \frac{E_{\rm rad,2015}}{1.8\times 10^{48} \, \erg} \right) \\
&
\left( \frac{R_2}{4 \rmRodot} \right)
\left( \frac{M_2}{10 \rmModot} \right)^{-1}
\rmModot.
\end{split}
\label{eq:macc2015}
\end{equation}
Taking $\beta=0.3$ can lower this estimate to $\approx 1.3 \rmModot$.
However, \cite{Thoneetal2016} estimate the total radiated energy of the 2015 eruption to be $\approx 1.8\times 10^{49} \erg$.
The HAPI model cannot account for such radiated energy in this system.

The estimate for $M_{\rm acc,2015}$ is similar to the estimate of $3.7 \rmModot$ for the mass accreted onto the companion of $\eta$ Car during the GE \citep{KashiSoker2010a}.
Though in the case of $\eta$ Car the eruption lasted for about twenty years, most of the accretion probably occurred very close to 2--4 periastron
passages, making the accretion time in the order of a week or two.
However, the 2015  event was short, $\approx 0.05 \yr$, and the implied accretion rate is huge, $\approx 4 \rmModot$.
We therefore regards the HAPI model less likely for the 2015 event.
The 2015 event is more likely a true supernova or a violent merger event. We next consider a violent head-on (or almost head-on) merger.

Adopting the accretion model for the 2015 eruption of SN~2015bh would need a strong stretching of the parameters, in particular a much more massive MS companion.
Instead, it is easier to account for the energy of the 2015 eruption if it came from a merger of the two stars.
The merger is actually an almost head-on collision of the two stars that were in a very high eccentric orbit before merger.
It is different from the onset of a CE phase. In the head-on collision case the gas-ejection morphology will be highly non-spherical.
If the secondary star is completely stopped as it hit the primary envelope in the head-on collision, then the energy that is released is
\begin{equation}
E_{\rm merger} \approx 3 \times 10^{49}  %1.9
\left( \frac{M_1}{50 \rmModot} \right)
\left( \frac{M_2}{10 \rmModot} \right)
\left( \frac{R_1}{50 \rmRodot} \right)^{-1}
\erg.
\label{eq:merger}
\end{equation}
A fraction of $10$ per cent will be enough to account for the radiated energy deduced by \cite{Ofeketal2016}, and a fraction of $\approx 50$ per cent is needed for the radiated energy calculated by \cite{Thoneetal2016}.
Namely, if during the very short time of the collision, a fraction of the orbital period, the velocity of the secondary star is reduced by $\approx 5-30$ per cent, the released energy can account for the energy of the outburst.
In the case of a merger-burst event, no further outbursts will take place.

% ==========================================================
\section{SUMMARY AND DISCUSSION}
 \label{sec:summary}
% ==========================================================

We discussed two recent ILOTs in the context of the binary model.
In section \ref{sec:M31LRN} we discussed the ILOT M31LRN~2015. We critically studied the model proposed by \cite{MacLeodetal2016}, according to which a low mass MS star entered a common envelope phase with a giant star of radius $\approx 35 \rmRodot$. The kinetic energy of the ejected gas results from the spiraling-in process of the companion into the envelope, and the radiation comes from the recombination of the ejected mass. We found severe problems with that model.
We suggested instead that M31LRN~2015 was an ILOT powered by a MS companion accreting from a giant, i.e., an ILRT type of ILOT (see section \ref{sec:intro} for terminology). The accretion is through an accretion disc that launches jets.
The jets carry most of the energy, such that the accreting star itself does not radiate much above its Eddington luminosity \citep{Shiberetal2016}.
The source of the radiated energy is the radiated energy by the disk and the conversion of kinetic energy to radiation when the jets collide with the material that was ejected a short time earlier.
In many cases accretion might be more efficient than recombination in powering ILOTs \citep{Soker2016NewA}. In the case of the ILOT M31LRN~2015 the companion needs to accrete a mass of $\approx 0.04 \rmModot$ (Equation \ref{eq:macc}).

In section \ref{sec:PTF13efv} we discussed the ILOT SN~2015bh (SNHunt~275) that underwent two eruptions, one in 2013 and another in 2015.
We raise the possibility that the 2015 outburst might it was a head-on collision, though it was more likely a terminal supernova explosion.
We examined the super-Eddington single star model as proposed by \cite{Ofeketal2016} for this ILOT. We found that the mass ejected according to the super-Eddington single-star model in the 2013 outburst is insufficient to account for the radius of the photosphere at $\gtrsim 10^{14} \cm$.
Here as well we suggested the HAPI-jets model, where the 2013 eruption was powered by accretion onto a companion in an eccentric orbit.
In case that the 2015 outburst was not a terminal supernova explosion, we speculated that the very energetic outburst might have been a head-on merger event of two massive stars (Equation \ref{eq:merger}).
Overall, with the presently available observations we cannot tell conclusively whether the 2015 was
a SN explosion, a terminal merger event, or an accretion event that left the binary system intact.

Over all, the binary model for ILOTs can account for different types of outbursts, but one should be careful in identifying the exact process.
Namely, which of the following processes takes place in each case (few of them can occur simultaneously): CEE, GEE, mass transfer, merger, and jet launching.

%\vspace{0.1cm}
% ==========================================================
\section*{Acknowledgments}
% ==========================================================
We thank Morgan MacLeod for helpful discussions about M31LRN~2015, Eran Ofek for clarifications about the observations of SNHunt~275,
and Nir Shaviv for discussions on the super-Eddington model.
We thank an anonymous referee for helpful comments.
%\vspace*{0.1cm}

\label{lastpage}
\end{document}